\documentclass[conference]{IEEEtran}
\IEEEoverridecommandlockouts
\usepackage{cite}
\usepackage{amsmath,amssymb,amsfonts}
\usepackage{algorithmic}
\usepackage{graphicx}
\usepackage{textcomp}
\usepackage{xcolor}
\usepackage{hyperref}
\usepackage{subfig}
\def\BibTeX{{\rm B\kern-.05em{\sc i\kern-.025em b}\kern-.08em
    T\kern-.1667em\lower.7ex\hbox{E}\kern-.125emX}}
\begin{document}

\title{Image classification using quantum inference on the D-Wave 2X}
\author{\IEEEauthorblockN{Nga T. T. Nguyen}
\IEEEauthorblockA{\textit{Information Sciences} \\
\textit{Los Alamos National Laboratory}\\
Los Alamos, New Mexico 87545, USA \\
}
\and
\IEEEauthorblockN{Garrett T. Kenyon}
\IEEEauthorblockA{\textit{Information Sciences} \\
\textit{Los Alamos National Laboratory \&}\\
\textit{New Mexico Consortium}\\
Los Alamos, New Mexico 87545, USA \\
gkenyon@lanl.gov}
}

\maketitle

\begin{abstract}
	We use a quantum annealing D-Wave 2X computer to obtain solutions to NP-hard sparse coding problems.  To reduce the dimensionality of the sparse coding problem to fit on the quantum D-Wave 2X hardware, we passed downsampled MNIST images through a bottleneck autoencoder.  To establish a benchmark for classification performance on this reduced dimensional data set, we built two deep convolutional neural networks (DCNNs).  The first DCNN used an AlexNet-like architecture and the second a state-of-the-art residual network (RESNET) model, both implemented in TensorFlow.  The two DCNs yielded classification scores of $94.54 \pm 0.7 \%$ and $98.8 \pm 0.1 \%$, respectively.  As a control, we showed that both DCNN architectures produced near-state-of-the-art classification performance $(\sim 99\%)$ on the original MNIST images.  To obtain a set of optimized features for inferring sparse representations of the reduced dimensional MNIST dataset, we imprinted on a random set of $47$ image patches followed by an off-line unsupervised learning algorithm using stochastic gradient descent to optimize for sparse coding.  Our single-layer of sparse coding matched the stride and patch size of the first convolutional layer of the AlexNet-like DCNN and contained $47$ fully-connected features, $47$ being the maximum number of dictionary elements that could be embedded onto the D-Wave $2$X hardware.  When the sparse representations inferred by the D-Wave $2$X were passed to a linear support vector machine, we obtained a classification score of $95.68\%$.  We found that the classification performance supported by quantum inference was maximal at an optimal level of sparsity  corresponding to a critical value of the sparsity/reconstruction error trade-off parameter that previous work has associated with a second order phase transition, an observation supported by a free energy analysis of D-Wave energy states.  We mimicked a transfer learning protocol by feeding the D-Wave representations into a multilayer perceptron (MLP), yielding $98.48\%$ classification performance.  The classification performance supported by a single-layer of quantum inference was superior to that supported by a classical matching pursuit algorithm set to the same level of sparsity. Whereas the classification performance of both DCNNs declined as the number of training examples was reduced, the classification performance supported by quantum inference was insensitive to the number of training examples.   We thus conclude that quantum inference supports classification of reduced dimensional MNIST images exceeding that of a size-matched AlexNet-like DCNN and nearly equivalent to a state-of-the-art RESNET DCNN.
\end{abstract}


\section{Introduction} \label{sec:introduction}
Deep learning has yielded impressive advances across a variety of machine learning tasks such as Alpha Go Zero\cite{alphagozero} and the ImageNet challenge\cite{imagenet}.
However, deep neural networks can be spoofed by adversarial examples\cite{goodfellow}, possibly due to an underlying reliance on low-level image statistics\cite{Jo}.  Moreover, the ability of deep neural networks to learn by analogy\cite{Ricci} as well as the importance of network depth in constructing abstract semantically-meaningful representations\cite{Tishby} has been called into question.
Unsupervised learning, on the other hand, particularly Boltzmann Machines\cite{Hinton} and sparse coding paradigms\cite{Olshausen_1996}, seek to learn joint distributions or causes directly from  unlabeled data  and thus may be less prone to some of the issues that have plagued task-specific deep learning approaches.
For example, features optimized in an entirely unsupervised manner for sparse coding can nonetheless support performance on image classification tasks that is only slightly below that achieved by standard deep neural network architectures such as AlexNet\cite{Coates}.
Sparse inference in these examples, however, relies on a convex approximation to the desired solution.
Optimal sparse inference, especially when derived from a highly overcomplete dictionary, is NP-hard, making it difficult to fully assess the potential of such approaches with existing algorithms.
Quantum computers offer a possible strategy for rapidly obtaining good solutions to NP-hard sparse inference problems.
Previous research\cite{NLK} demonstrated that models of sparse inference based on lateral inhibition between binary neurons can be directly implemented on the D-Wave 2X Quantum Annealing Computer.   
Here, we show that on a suitably low-dimensional problem that fits on available quantum annealing architectures, a single layer of features optimized in an unsupervised manner for sparse inference followed by a multi-layer perceptron (MLP) nearly matches the classification performance of a state-of-the-art RESNET deep convolutional neural network (DCNN) trained in a fully supervised manner for the specific task in question.  

Our paper is organized as follows.  Sec.~\ref{sec:sparse-coding}, Sec.~\ref{sec:quantum-inference}, and Sec.~\ref{sec:inputs} introduce concepts, methodology, and reduced dimensional images for mapping a binary sparse coding classification problem onto the D-Wave 2X machine.  Classification based on D-Wave generated sparse representations are compared with with the state-of-the-art DCNN classifiers in Sec.~\ref{sec:Uclassification} and Sec.~\ref{sec:classification}.  Evidence for a phase-transition based on a free energy analysis of solutions from the D-Wave quantum annealer is presented in Sec.~\ref{sec:PT}. 

\section{Sparse-coding} \label{sec:sparse-coding}
The hypothesis that neurons encode stimuli by inferring sparse representations explains many of the response properties of simple cells in the mammalian primary visual cortex\cite{Olshausen_1996,Olshausen_1997}.  Given an overcomplete, non-orthonormal basis $\{  \phi_i \}$, inferring a sparse representation involves finding the minimal set of non-zero activation coefficients $\boldsymbol{a}$ that accurately reconstruct a given input signal $\boldsymbol{X}$, corresponding to a minimum of the following energy function: 
\begin{equation}
	E(\boldsymbol{X}, \boldsymbol{\phi}, \boldsymbol{a}) = \min\limits_{ \{\boldsymbol{a}\} } [  \, \frac{1}{2}  ||  \boldsymbol{X} - \boldsymbol{\phi} \boldsymbol{a} ||^2	+	\lambda || \boldsymbol{a} ||_0 \, ]
	\label{eq:H_SC}
\end{equation}
where $\lambda$ is a trade-off parameter that determines the balance between reconstruction error of the original input image $\boldsymbol{X}$ and the number of non-zero (sparse) activation coefficients.  A larger $\lambda$ encourages sparser  solutions.  We define $\gamma = \frac{\text{rank}(\boldsymbol{a})}{\text{rank}(\boldsymbol{X})}$ as the overcompleteness of the basis $\{  \phi_i \}$.  $\gamma$ typically is chosen such that $\gamma \gg 1$ which means that in general there will exist many solutions which achieve the same  $||\boldsymbol{X} - \boldsymbol{\phi} \boldsymbol{a} ||^2$ and our task is to find the sparsest one.  The energy function Eq.~(\ref{eq:H_SC}) is non-convex and contains multiple local minima, so that finding a sparse representation falls into an NP-hard complexity class of decision problems\cite{Natarajan_1995}.

\section{Quantum inference} \label{sec:quantum-inference}
\subsection{Mapping Sparse coding problem on D-Wave 2X}
The D-Wave 2X\cite{DWAVE} consists of 1152 quantum bits (qubits) arranged into $12$x$12$ unit cells, forming a Chimera structure with dimensions $12$x$12$x$8$. Sparse interactions between qubits are restricted to the $16$ connections within a unit cell and the $16$ connections between nearest-neighboring unit cells\cite{DWAVE}. 

In detail, each unit cell contains $4$ qubits aligned along a horizontal axis and $4$ qubits aligned vertically.  Within a unit cell, the $4$ qubits of a given orientation can only can interact with the $4$ qubits with the opposite orientation (details see\cite{NLK}). Between unit cells, interactions are only allowed between nearest-neighbors and even between nearest-neighbors the allowed connections are  restricted according to relative orientation.  A vertically (horizontally) oriented qubit can only connect to the two vertically (horizontally) oriented qubits at the same relative position in the nearest-neighboring unit cells immediately above (left) and below (right) along one column of the $12$x$12$ grid.  Thereby, in a chimera graph, one qubit can interact with \textit{at most} $6$ other qubits. 

The D-Wave 2X\cite{DWAVE} finds optimal solutions to a (discrete) Ising system consisting of $N_q$ binary variables via quantum annealing.  Such $N_q$-body systems can be described by the following classical Hamiltonian:
\begin{equation}
	H(\boldsymbol{h}, \boldsymbol{Q}, \boldsymbol{a}) = \sum_i^{N_q} {h_i a_i} + \sum_{i<j}^{N_q} Q_{ij} a_i  a_j
	\label{eq:H_Dwave}
\end{equation}
with binary activation coefficients $a_i = \{ 0, 1\} \, \forall{i} \in (1, 2, 3, ..., N_q) $.  This objective function defines a quadratic unconstrained binary optimization (QUBO) problem. We cast our sparse coding problem, Eq.~(\ref{eq:H_SC}), into QUBO form, Eq.~(\ref{eq:H_Dwave}), by the transformations\cite{LANL}: 
\begin{eqnarray}
	h_i 		&=& (-\boldsymbol{\phi}^{T}  \boldsymbol{X} + (\lambda + \frac{1}{2}))_i,  \nonumber \\
	Q_{ij} 	&=& (\boldsymbol{\phi}^{T} \boldsymbol{\phi})_{ij}.
	\label{eq:hQ}
\end{eqnarray}
In Eq.~(\ref{eq:hQ}), the bias term $\boldsymbol h$ (elements $h_i$) in the Ising model is proportional to the weighted input $\boldsymbol{\phi}^{T}  \boldsymbol{X}$ while the coupling term $\boldsymbol Q$ (elements $Q_{ij}$) corresponds to lateral competition (see also \cite{Rozell_2008}) between qubits given by the interaction matrix $\boldsymbol{\phi}^{T} \boldsymbol{\phi}$ [with self-interaction excluded and $\boldsymbol Q$ being symmetric i.e. $Q_{ij} = Q_{ji} \,\, \forall i \ne j$, see Eq.~(\ref{eq:H_Dwave})].  The trade-off parameter $\lambda$ corresponds to a uniform applied field that biases each qubit to be in the zero state.

From Eq.~(\ref{eq:hQ}), we infer a sparse representation to Eq.~(\ref{eq:H_SC}) on D-Wave 2X hardware by associating each neuron with a single feature $\phi_i$, represented as a binary logical qubit, with logical qubits embedded on the D-Wave physical chimera as follows. 

\subsection{Embedding} 
Despite the sparsity of physical connections on the D-Wave, it is nonetheless possible to construct graphs with arbitrarily dense connectivity by employing ``embedding" techniques.  Embedding works by chaining together physical qubits so as to extend the effective connectivity but at the cost of reducing the total number of available logical qubits.  Because logical qubits do not need to follow the connection rules that physical qubits do, it is possible to implement general QUBO problems with arbitrarily dense connectivity on the sparsely connected D-Wave chimera.  The D-Wave API provides a heuristic algorithm that searches for an optimal embedding that minimizes the number of physical qubits that are chained together. 

The exact mapping of a spin glass problem onto the physical D-Wave 2X chimera, including defects, can typically contain approximately $N_q \sim 1000$ spins (qubits) with $> 3000$ local spin-spin interactions.  In contrast, embedding an arbitrary QUBO problem onto the same 2X chimera typically allows no more than $N_q \sim 47$ nodes (logical qubits) but these nodes may be fully connected.  Thus, embedding effectively trades qubits for connectivity. 

Because a logical qubit is assigned only two possibilities $0$ and $1$, each neuron is treated as a ``quantum object'' bearing two possibilities: firing with maximum activation with coefficient $1$ and silent with coefficient $0$.  Because each neuron is a quantum object, the state of any neuron is described in general by a superposition of $1$ and $0$, in which the neuron is both active and non-active at the same time, a logical impossibility for any classical system.  Due to this quantum superposition, the annealing processing in D-Wave allows to explore the entire energy landscape at once.  

\section{D-Wave inputs: reduced dimensional images} \label{sec:inputs}
 %
\begin{figure}
  \vspace{0.0in}
  \hspace{0.in}
  \begin{center}
  	\includegraphics[width=0.5\textwidth]{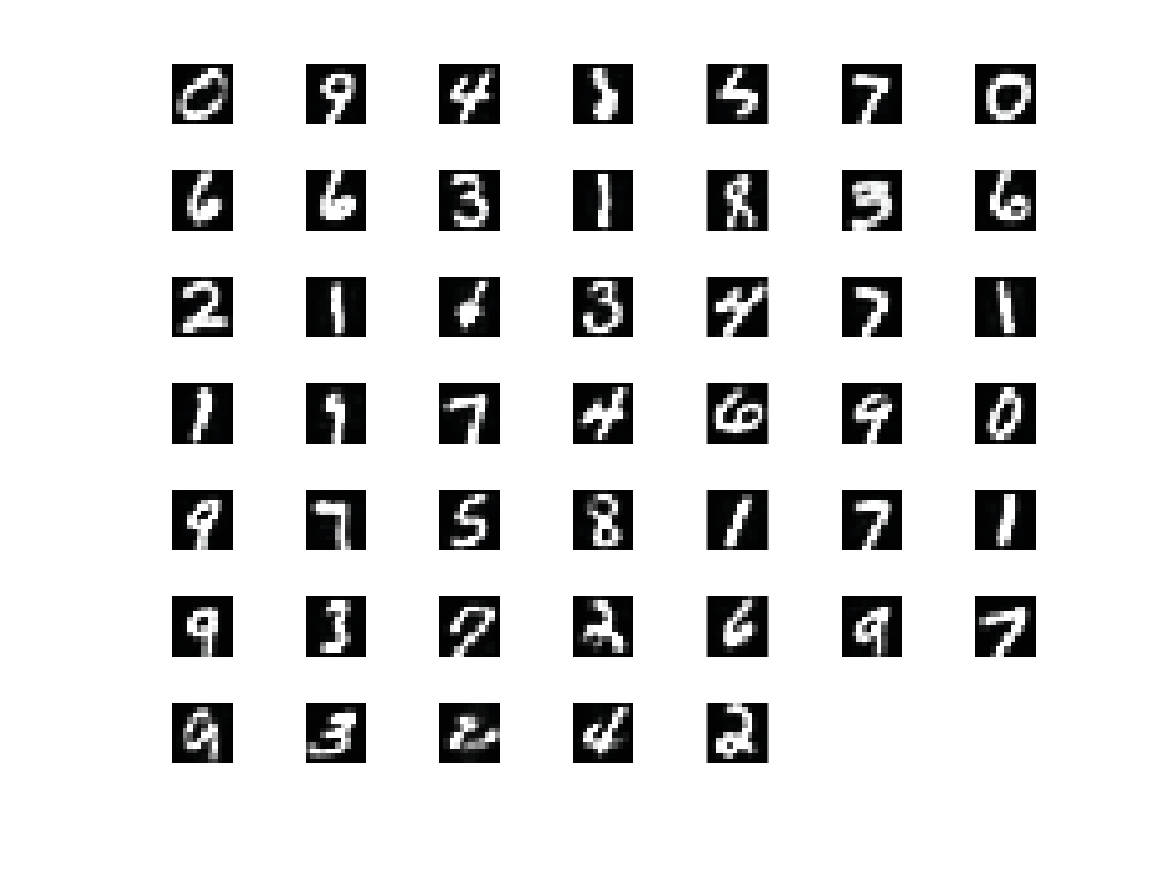}
  \end{center}
  \vspace{-0.18in}
  \caption{$47$ $12$x$12$ randomly selected reduced dimensional images reconstructed through a bottleneck autoencoder, yielding an undercomplete dictionary of 47 features for generating sparse representations of the reduced dimensional MNIST images.  To construct an overcomplete dictionary, we randomly selected a set of $47$ $6$x$6$ patches and divided each reduced dimensional MNIST image into overlapping $6$x$6$ patches with a stride of 2 and separately generated a sparse representation of each $6$x$6$ patch. The resulting $47$ imprinted features, whether $12$x$12$ or $6$x$6$, were further refined using unsupervised SGD to optimize for sparse reconstruction.}
  \vspace{0.2cm}
  \label{fig:AEfeatures}
\end{figure}
\begin{figure}
  \vspace{0.0in}
  \hspace{0.in}
  \begin{center}
  	\includegraphics[width=0.45\textwidth]{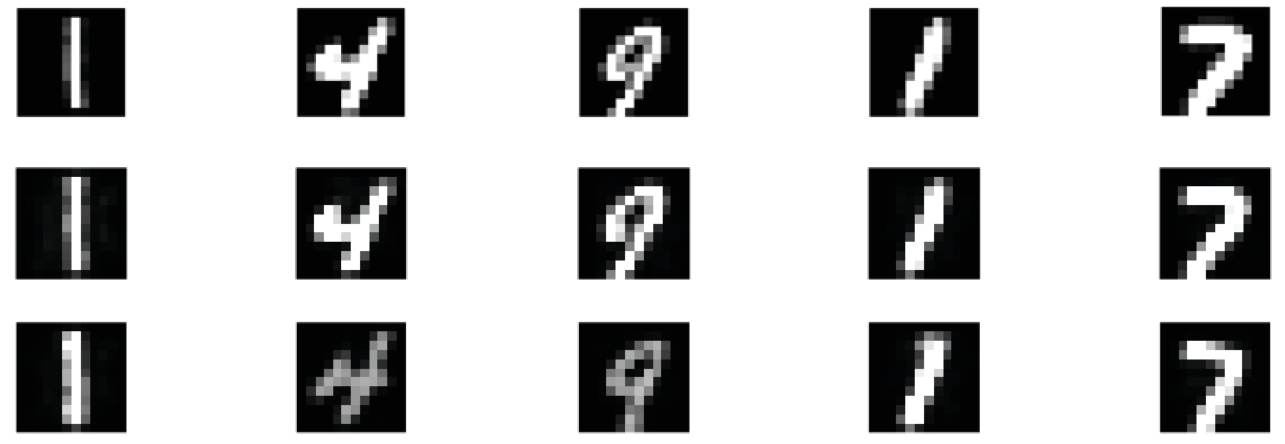}
  \end{center}
  \vspace{0.00in}
  \caption{ Top: Original MNIST images downsampled to $12$x$12$, Middle: Reconstructed images from bottleneck autoencoder, Bottom: D-Wave reconstruction using undercomplete dictionary based on randomly selected imprinted features.}
  \vspace{0.2cm}
  \label{fig:recon}
\end{figure}
We first downsampled original $28$x$28$ MNIST images to $12$x$12$  and trained a bottleneck autoencoder (using TensorFlow) that consisted of $47$ neurons at its narrowest point.  The resulting autoencoder generated images, examples of which are shown in Fig.~\ref{fig:AEfeatures}, constitute a database of images with greatly reduced dimensionality suitable for analysis by the D-Wave 2X.  Constructing a dictionary based on a randomly chosen set of $47$ $12$x$12$ reduced dimensional MNIST images results in undercomplete basis $\gamma = 47/144$ ($\approx 0.33$) $\ll 1$. We used this undercomplete basis to demonstrate that it was possible to obtain sparse representations of reduced dimensional MNIST images on the D-Wave 2X (Fig.~\ref{fig:recon}).  Using an undercomplete basis however reduces the advantages of sparse inference as the resulting representations are unlikely to be sparse.

\section{Classification performance: Undercomplete dictionary} \label{sec:Uclassification}
\subsection{Image classification}
Due to computational limitations, it was not possible to perform all of the analyses presented in this paper using an overcomplete dictionary for generating sparse representations.  We therefore first present benchmark performance using an undercomplete dictionary.  Benchmark results obtained using an undercomplete dictionary  helps to illustrate the advantages of sparse inference on the D-Wave quantum annealing machine when using an overcomplete basis.  

Recall that reduced dimensional images were obtained by first downsampling to $12$x$12$ and then reconstructing through a bottleneck autoencoder employing a hidden layer containing $47$ feature vectors, thereby ``compressing'' the $144$ pixels in the downsampled images to a $47$ dimensional manifold.  Thus, using a dictionary containing $47$ features might be considered complete, although nominally $\gamma = 47/144$.  

\begin{figure}%
    \centering
    \subfloat[]{{\includegraphics[width=6.cm]{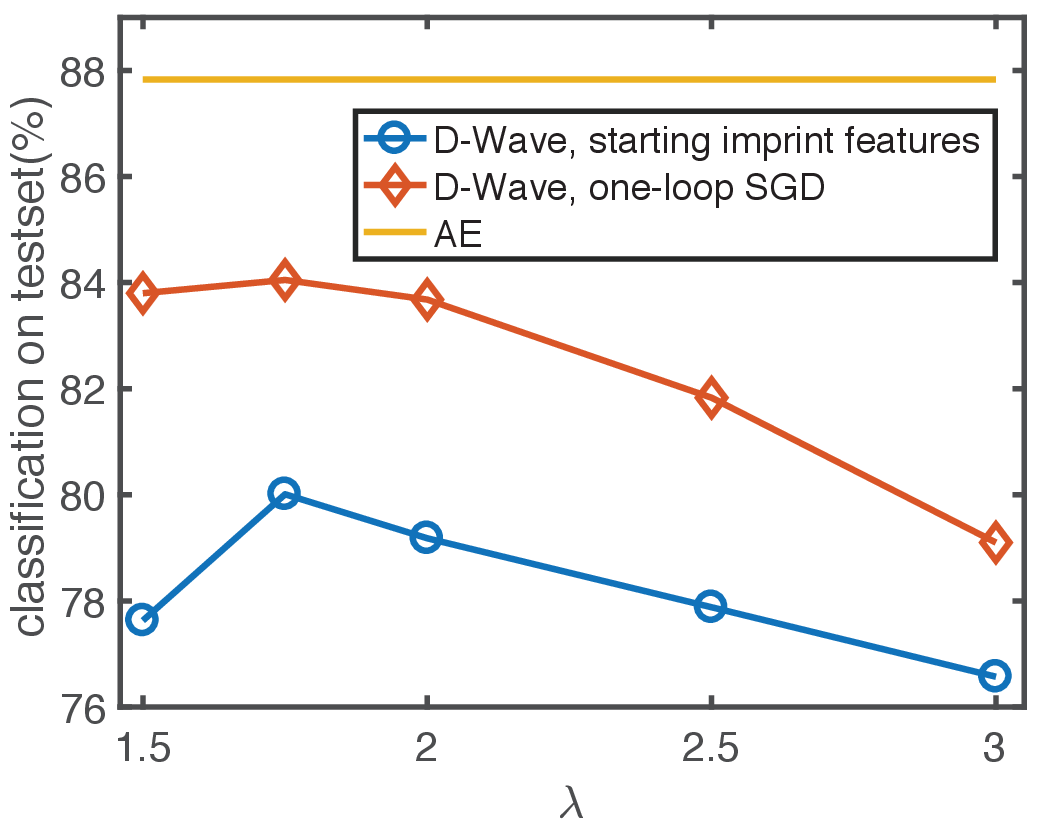} }}%
    \qquad
    \subfloat[]{{\includegraphics[width=6.cm]{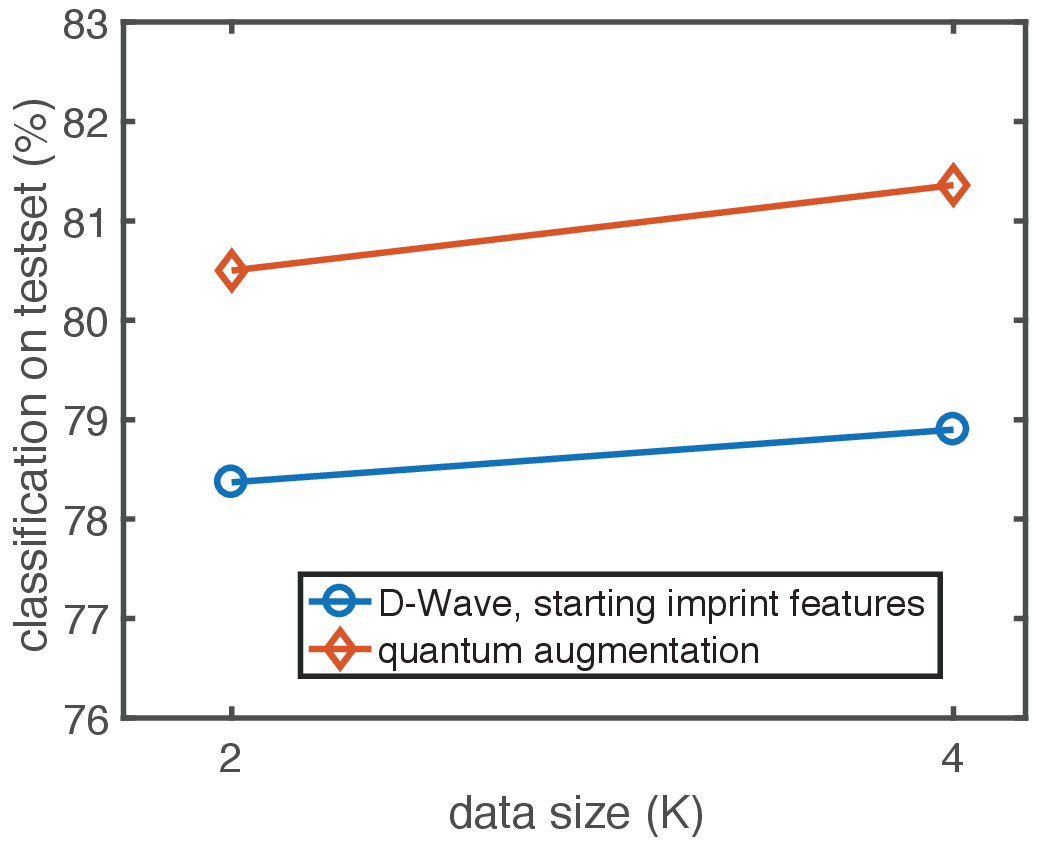} }}%
    \caption{(Color online) (a): Image classification obtained by feeding the sparse representations of the downsampled autoencoded MNIST images generated by the D-Wave 2X  before (blue curve) and after unsupervised feature optimization (red curve) to a linear SVM.  Dataset contains $55$ K images.  The classification result obtained using the autoencoder latent representations as input to a linear SVM was $87.83\%$ (yellow line).  (b): Image classification on up to $4,000$ sparse representations of downsampled autoencoded MNIST images generated by the D-Wave 2X using the lowest-energy state (blue circles) and quantum augmentation using the $30$ lowest energy level states (red diamonds).  Training and test sets were divided in a ratio of 5/1, respectively.  Dataset size is in units of $1,000$ images.}%
    \label{fig:classification12x12}%
\end{figure}

Fig.~\ref{fig:classification12x12}(a) shows classification results on reduced dimensional $12$x$12$ MNIST images using an undercomplete dictionary for sparse coding containing $47$ randomly selected imprinted features.  
The sparse representations generated by the D-Wave 2X for the complete set of $55$ K reduced dimensional MNIST images was passed to a linear SVM.  
As a function of the trade-off parameter $\lambda$, classification performance using an undercomplete basis peaked at approximately $80$\%.  
Here, sparsity was approximately $ \{ 16\%, 13\%, 12\%, 9\%,  5\% \}$  for $\lambda$ of $ \{ 1.5, 1.7, 2.0, 2.5, 3.0 \}$, respectively. 
By comparison, simply passing the latent representations from the hidden layer of the autoencoder to an SVM yielded a classification score nearly $88$\% [see Fig.~\ref{fig:classification12x12}(a), note that a bottleneck autoencoder is independent of $\lambda$].   Thus, when using an undercomplete basis, we find no advantage to generating sparse representations on the D-Wave 2X.

Randomly selected imprinted features are unlikely to be optimal for sparse coding, which likely reduced the classification performance using the resulting sparse representations. 
We therefore conducted unsupervised training to optimize our undercomplete basis for sparse coding. Our unsupervised training procedure used a local Hebbian rule with momentum for computing weight updates to optimize our initially imprinted features $\boldsymbol{\phi}$.  

The SGD loop required generating sparse representations on the D-Wave 2X for a randomly chosen mini-batch consisting of $10$K reduced dimensional MNIST images, from which we computed an average weight update for the undercomplete dictionary $\boldsymbol{\phi}$.  This cycle was repeated for approximately a few thousand steps until the loss function saturated.  After unsupervised optimization, we obtained an approximately  $4\%$ increase in the peak classification accuracy [red curve in  Fig.~\ref{fig:classification12x12}(a)].

\subsection{Quantum augmentation}
Quantum annealing on the D-Wave is able to return multiple sparse solutions corresponding to a set of good (local) minima given an input $\{\boldsymbol{h}, \boldsymbol{Q} \}$.  We examined whether these multiple sparse solutions could support improved image classification by treating these distinct representations analogously to standard image augmentation procedures.  Due to the cost of computation for this task, we present the classification scores on only $2$ K and $4$ K ``quantum augmentation'' images (with the ratio between training/test sets $5$/$1$) of the reduced dataset for the ``best'' $\lambda = 1.7$ obtained from Fig.~\ref{fig:classification12x12}(a).  We show in Fig.~\ref{fig:classification12x12}(b) that a quantum augmentation consisting of $30$ sparse solutions extracted from the $2$X hardware with an undercomplete dictionary gains about $\sim 2\%$ in classification in this case. 

\section{Classification performance: Overcomplete dictionary} \label{sec:classification}

The above classification results obtained via quantum inference used a single $1$x$1$x$47$ feature map fed directly into a linear SVM.  In this section, we generate sparse representations on the D-Wave 2X using an overcomplete dictionary by breaking the $12$x$12$ reduced-dimensional inputs into overlapping $6$x$6$ tiles.  
\subsection{Patch, stride}  
 
To construct an overcomplete dictionary (i.e. $\gamma > 1$), we represent each image as a set of overlapping patches. Sliding a $6$x$6$ patch throughout the entire image of total size $12$x$12$ with a stride $s=2$, we obtain a set of $4$x$4$ (= $16$) patches for each image.  The dictionary, which now consists of $47$ randomly selected $6$x$6$ patches, has overcompleteness of $\gamma = 47/36 = 1.31$.   Consequently, the feature map now increases from $1$x$1$x$47$ to $4$x$4$x$47$.    

\subsection{Evaluating classification performance}

\begin{figure}
  \vspace{0.0in}
  \hspace{0.in}
  \begin{center}
  	\includegraphics[width=0.4\textwidth]{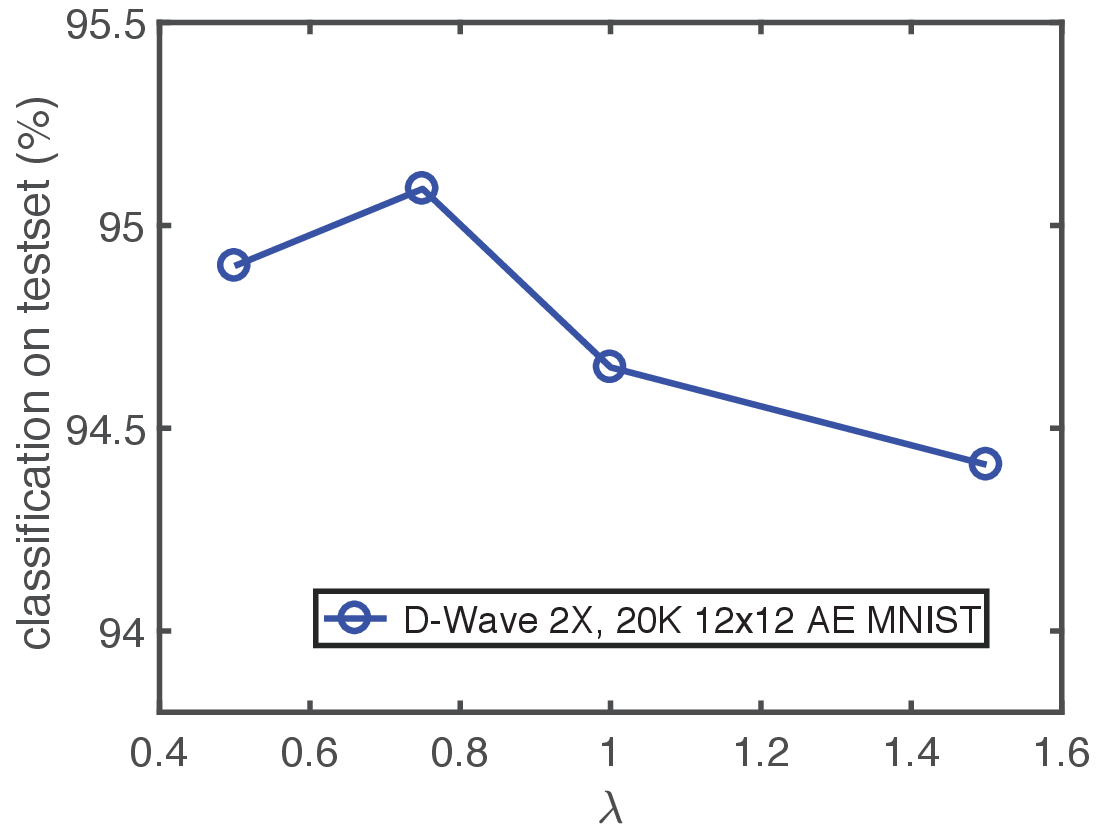}
  \end{center}
  \vspace{0.00in}
  \caption{(Color online) Image classification using an overcomplete dictionary applied separately to overlapping $6$x$6$ image patches on the D-Wave 2X for the first $20$ K images of our $12$x$12$ autoencoded MNIST dataset plotted as a function of $\lambda$.  Training and test sets were divided in a ratio of 5/1, respectively. This procedure established a best value of $\lambda$. }
  \vspace{0.2cm}
  \label{fig:20Kscores}
\end{figure}

To determine an optimal value for the trade-off parameter $\lambda$, we study the four different cases of $\lambda = \{ 0.5, 0.7, 1.0, 1.5 \}$.  This corresponds, respectively, to sparsity of $\{ 14\%, 12\%, 10\%,$ and $8\% \}$ on average.  For classification, we employed the ``feature extraction'' technique (see for example Ref.~\cite{Coates}) to obtain a sparse representation for each image by concatenating the binary sparse coefficients for each $6$x$6$ image patch into a $4$x$4$x$47$ array.  We used the LIBLINEAR package\cite{liblinear} to train an SVM classifier on this $10$-category image classification task using the combined sparse coefficients.  For the first $20$K images, because each $12$x$12$ image has $16$ $6$x$6$ overlapping tiles, there were $16$x$20$K ($=320$K) input Hamiltonians to solve using the D-Wave $2$X.  As a function of $\lambda$, we obtained the classification scores shown in Fig.~\ref{fig:20Kscores}.  By replacing the undercomplete dictionary applied to the entire image with an overcomplete dictionary applied to each image patch,  we improved our classification results based on quantum inference substantially.  We obtain a peak behavior for $\lambda_c = 0.7$ which translates into $\sim 12\%$ of sparsity, corresponding to an average of $5-6$ active neurons  to reconstruct each $6$x$6$ tile. \\

\begin{figure}
  \vspace{-1.in}
  \hspace{-0.2in}
  \begin{center}
  \includegraphics[width=0.5\textwidth]{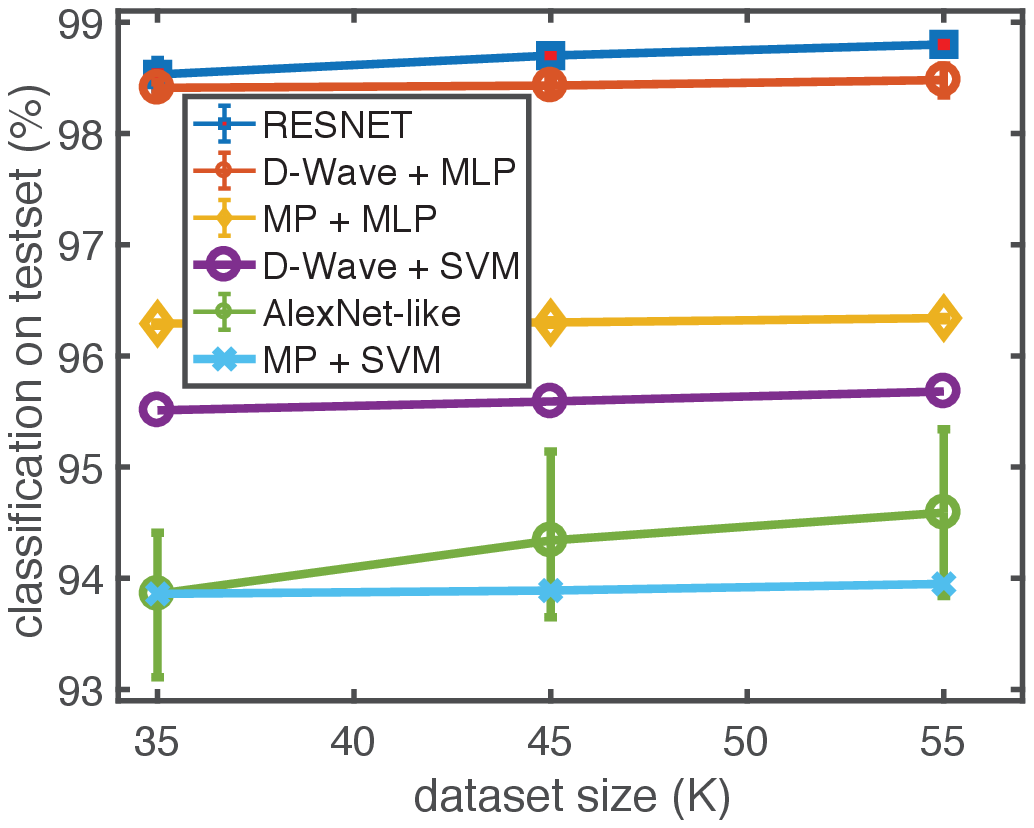}
  \end{center}
  \vspace{-1in}
  \caption{(Color online) Image classification on the D-Wave 2X, near state-of-the-art DCNN (AlexNet-like) built with TensorFlow, and matching pursuit for our reduced dimensional $12$x$12$ autoencoded MNIST dataset plotted as a function of dataset size.  Training and test sets in each case are in units of 1,000 images and divided in a ratio of 5/1, respectively.  We also show classification results using state-of-the-art RESNET for our customized MNIST images.  }
  \vspace{0.0cm}
  \label{fig:comparison}
\end{figure}

\textbf{Image classification: D-Wave 2X vs. DCNN}\\
As an initial benchmark, we started with an AlexNet-like DCNN architecture that approximately matched the size of the latent representations generated by the D-Wave 2x. The AlexNet-like DCNN possessed an initial convolutional layer with a stride of $2$ and $12$ features, yielding a set $12$ $4$x$4$ feature maps, with each being the same size as the feature maps generated by the D-Wave 2X.  A subsequent set of $16$ $5$x$5$ feature maps was followed by a final all-to-all layer using $47$ features, the same number of feature maps on the D-Wave 2x, which in turn fed into a SoftMAX classifier (cross-entropy loss).  
Fig.~\ref{fig:comparison} shows that quantum inference enabled superior classification performance compared to an AlexNet-like DCNN.  When applied to the full reduced dimensional MNIST dataset, the classification accuracy of the AlexNet-like DCNN was $94.54 \pm 0.7 \%$ (green connected circles with error bars).  The same AlexNet-like DCNN produced near-state-of-the-art results of $\sim 98.9\%$ on the original MNIST $28$x$28$ pixel dataset (not plotted), affirming the validity of our DCNN implementation.  Likewise, if we remove the autoencoder procedure and use the downsampled MNIST images to train the AlexNet-like DCNN model, we obtained an accuracy of $96.45\%$ (not plotted), consistent with the expectation that the autoencoder destroys some information.  Sparse representations generated by the D-Wave 2X, when fed to a linear SVM classifier\cite{liblinear}, produced a slightly higher classification score than the AlexNet-like DCNN, $95.68\%$ (purple circles) and exhibited almost no variation between reinitialized runs.  Additionally, classification by the AlexNet-like DCNN declined as the size the training set was reduced whereas classification based on quantum inference showed no such degradation.  Matching pursuit\cite{Mallat}, a classical technique for solving sparse coding problems, produced a lower classification score than did quantum inference when the former was forced to match the same sparsity as the  solutions generated by the D-Wave ($\sim 12-14\%$).  

We further examined a transfer learning procedure where the D-Wave sparse representations were fed into a multilayer perceptron (MLP), yielding a classification of $\sim 98.48\%$ (red circles).  Feeding the sparse representations produced by a classical matching pursuit algorithm into an MLP yielded a lower classification score of $\sim 96.34\%$ (yellow diamonds), suggesting that quantum inference may be superior to classical solutions.  The AlexNet-like DCNN was tailored to approximately match the size of the sparse representations generated by the D-Wave 2x.  We also tested a state-of-the-art  (RESNET) DCNN and obtained a classification of $\sim 98.8\%$ (blue squares) on the reduced MNIST database.  While RESNET yielded a slightly higher classification score using all of the available training data, this advantage disappeared as the amount of training data was reduced.  Using a reduced dimensional dataset that fits onto the D-Wave 2X hardware reveals the benefits of quantum inference for image classification.

\section{Phase-transition in a Sparse-coding model}\label{sec:PT}
	Previous work\cite{CCK} has revealed the existence of a $2$nd order phase transition in the sparse representations generated by Locally Competitive Algorithm (LCA) implemented using continuously-valued neurons simulated on a conventional digital computer.   Here, we search of evidence of a phase transition in the binary sparse coding problem solved on the D-Wave 2X by assuming a thermal Boltzmann distribution for the spectrum of energy states generated by different annealing runs.   

\begin{figure}%
    \centering
    \subfloat[]{{\includegraphics[width=6.cm]{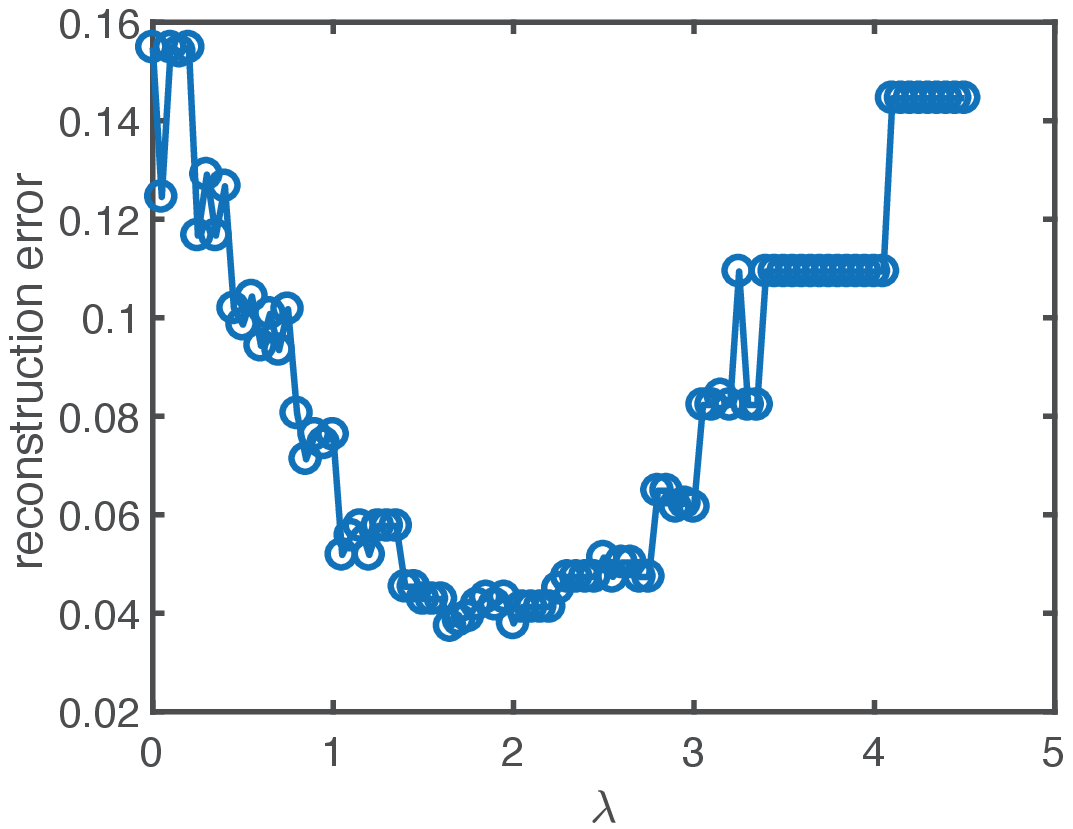} }}%
    \qquad
    \subfloat[]{{\includegraphics[width=6.cm]{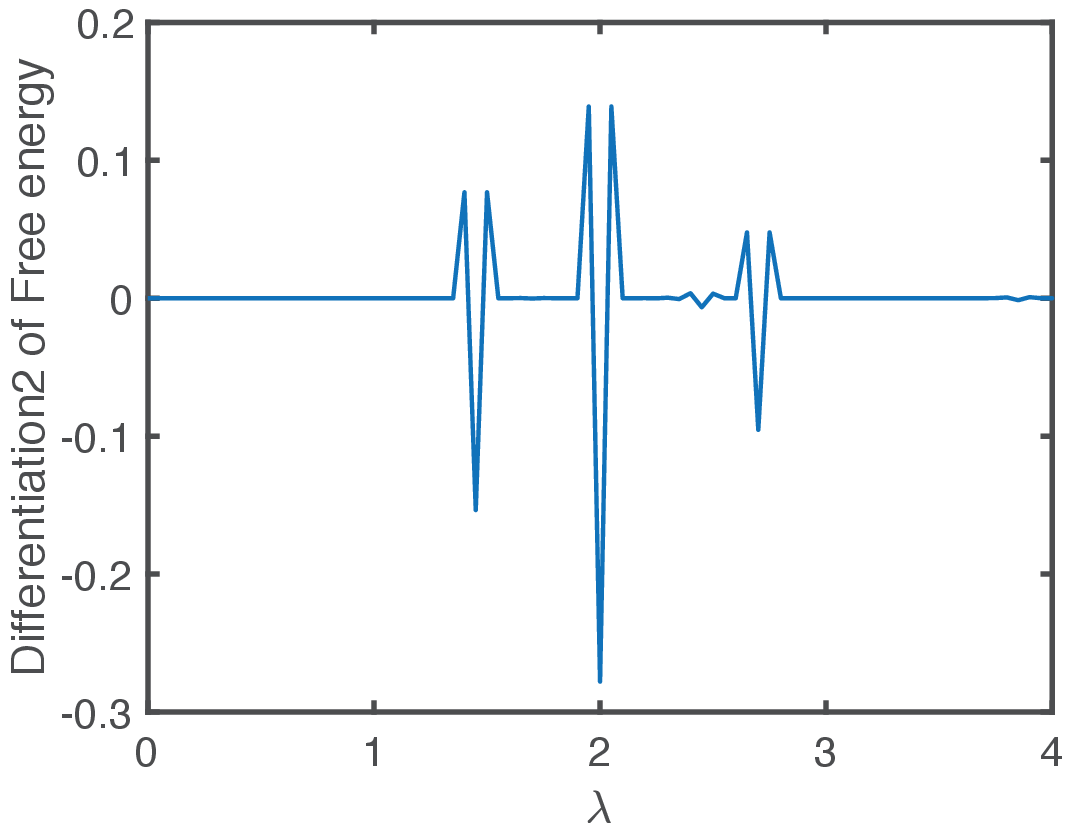} }}%
    \caption{(Color online) Top: Reconstruction error in a denoising sparse coding system for one autoencoded-$12$x$12$ image plotted as a function of sparsity threshold.  Bottom: Second derivative of the free energy with respect to $\lambda$ of the system described in the top plot at temperature $k_B T = 10^{-2}$ (energy unit). }%
    \label{fig:2ndPT}%
\end{figure}
We consider the sparse representation of one randomly-chosen  $12$x$12$ reduced dimensional MNIST image.   To explore statistical properties of the sparse representations inferred on a D-Wave hardware, we treat the trade-off parameter $\lambda$ as an ``effective'' (magnetic) field, where $\lambda$ ``couples'' to an Ising model through the qubit-field interaction in $\boldsymbol{h}$.  Next, the energy spectrum $\{ E_i \}$ is obtained using different annealing runs, corresponding to parameter deviations in $\{ \boldsymbol h, \, \boldsymbol Q \}$.  We executed multiple runs and obtained over ten thousand sparse solutions along with the corresponding energy spectrum for a particular reduced dimensional MNIST image.  We used that energy spectrum  to examine the $2$nd derivative of free energy, defined as $F  = -\frac{1}{\beta}  \log Z$, with respect to the ``field'' $\lambda$ in  Fig.~\ref{fig:2ndPT}. Here, $Z = \sum_{i=1}^{N_s} { e^{-\beta E_i} }$ is defined as partition function (\cite{Pathria}), $\beta=1/k_B T$ with $k_B$ is Boltzmann constant and  $T$ the temperature of the embedded sparse coding model.  $N_s$ is the total number of $N_q$-body states obtained in this case by solving Eqs.~ (\ref{eq:H_SC}) and ~(\ref{eq:hQ}) to yield the corresponding energy set $\{ E_i \}$.   The reconstruction error calculated using the lowest-energy level as a function of sparsity $\lambda$, presented in Fig.~\ref{fig:2ndPT}(a), shows that there exists a critical $\lambda_c \approx 2$ that yields the best denoising result (i.e. yields the most accurate reconstruction of the input).   This optimal behavior coincides with sharp discontinuity around $\lambda \approx 2$ in the second derivative of $ F $ with respect to $\lambda$ in Fig.~\ref{fig:2ndPT}(b).  The critical point is consistent with the optimal values of $\lambda$ found above.     

 
\section{Conclusions}
We have explored classification performance on dimensionally-reduced binary MNIST images using sparse representations generated by the D-Wave 2X quantum annealing computer at a putative critical value of sparsity.  Given the limited number of qubits available on the D-Wave 2X, we first used a bottleneck autoencoder to reduce the intrinsic dimensionality of MNIST images, which originally consisted of 28x28=784 binary values.   A bottleneck autoencoder is trained in a standard manner using backprop and SGD to reconstruct, as accurately as possible, input images that have been forced through a narrow “waist”, corresponding to a latent representation whose dimensionality (number of neurons in the layer comprising the waist of the hourglass) is less than the dimensionality of the original image.  

To investigate how quantum inference might contribute to classification performance, we added lateral inhibition between 47 features obtained by randomly choosing (imprinting) from a subset of reduced $12$x$12$-pixel reconstructed MNIST images, allowing us to generate a sparse binary representation for each reduced-dimensional image via quantum inference.   Although the reduced database itself was constructed to live on a $47$-dimensional manifold, local regions of the manifold are likely to be characterized by an even smaller number of basis elements.  In principle, this reduced basis can be inferred via quantum inference.  As a target benchmark, we trained standard deep neural network classifiers, implemented in TensorFlow, on our reduced dimensional database.  A size matched deep learning architecture yielded a classification score of $94.54 \pm 0.7 \%$ compared to $98.48\%$ obtained by feeding sparse representations inferred on the D-Wave 2X into an MLP.  Classification of state-of-the-art model (RESNET) was only slightly higher.  The RESNET model however contains orders of magnitude more neurons and free parameters compared to quantum inference.  We showed that quantum inference on the D-Wave $2$X, using approximately $47$ binary neurons (logical qubits) and only a single layer of convolutional kernels optimized for sparse reconstruction using an unsupervised training procedure, could match the classification performance obtained by a state-of-the-art deep neural network as the size of the training set was reduced. 

Quantum inference on the D-Wave hardware allowed us to examine a postulated thermodynamical phase-transition in a sparse coding system  by analyzing a large number of energy states obeying a Boltzmann distribution.  We found a critical sparsity $\lambda_c$ as the minimal point in the reconstruction error and this optimal $\lambda_c$ that appears as sharp discontinuity in the $2$nd differentiation of the thermodynamically averaged free energy produces the smallest image classification error. 

\section*{Acknowledgment}
We acknowledge Marcus Daniels and Jack Raymond for helpful discussions. 

\end{document}